\documentclass[letterpaper, 10 pt, conference]{ieeeconf}  % Comment this line out if you need a4paper

\IEEEoverridecommandlockouts                              % This command is only needed if 
\usepackage{cite}
\usepackage{amsmath,amssymb,amsfonts}
\usepackage{lipsum}% For this example
\usepackage[dvipsnames]{xcolor}
\usepackage{algorithmic}
\usepackage{makecell}
\usepackage[linesnumbered,ruled,vlined]{algorithm2e}
\SetKwInput{KwInput}{Input}                % Set the Input
\SetKwInput{KwOutput}{Output}              % set the Output
\SetKwInOut{Parameter}{parameter}
\usepackage{alphabeta} 
\usepackage[utf8]{inputenc}
\usepackage{commath} % Nemt at lave differntieringssymboler. Ex: \dif t. Ex: \od{f}{x}. Ex: \pd
\usepackage{nccmath} % Til lidt mindre matrixer
\usepackage{pifont}
\usepackage{optidef}
\usepackage{array}
\usepackage[arrowdel]{physics}

\usepackage[position=bottom,caption=false]{subfig}
\usepackage{multirow}                		% Fletning af raekker og kolonner (\multicolumn og \multirow)
\usepackage{colortbl} 						% Farver i tabeller (fx \columncolor, \rowcolor og \cellcolor)

\usepackage{tabularx}
\usepackage{siunitx}						% Flot og konsistent praesentation af tal og enheder med \si{enhed} og \SI{tal}{enhed}
\usepackage{nameref}
\let\labelindent\relax
\usepackage{enumitem}
\setitemize[0]{leftmargin=*}
\usepackage{hyperref}
\usepackage{tikz}
\usepackage{pgfplots} 
\usetikzlibrary{external}
\tikzsetexternalprefix{Figures/}
\usetikzlibrary{calc,patterns,angles,quotes,backgrounds,shapes,arrows,positioning}
\pgfplotsset{compat=1.6}
\usepgfplotslibrary{groupplots}
\usepgfplotslibrary{dateplot}
\usetikzlibrary{decorations.markings,shapes.arrows}
\usetikzlibrary{calc,patterns,angles,quotes,backgrounds,shapes,arrows,positioning}
\usepgfplotslibrary{colorbrewer}
\usetikzlibrary{colorbrewer}

\definecolor{matBlue}{rgb}{0,0.447,0.741}
\definecolor{matRed}{rgb}{.85,.325,.098}
\definecolor{matYellow}{rgb}{0.92,.694,.125}
\definecolor{matPurple}{rgb}{0.494,0.184,0.556}
\definecolor{matGreen}{rgb}{0.466,0.674,0.1880}
\definecolor{matBlueLight}{rgb}{0.3010, 0.7450, 0.9330}
\definecolor{matRedDark}{rgb}{0.6350 0.0780 0.1840}

\pgfplotscreateplotcyclelist{matlab}{
    matBlue, matRed, matYellow, matPurple, matGreen, matBlueLight, matRedDark
}

\pgfkeys{/pgf/number format/.cd,1000 sep={}} % Thousand separator
\pgfplotsset{every axis/.append style={xticklabel style={/pgf/number format/fixed,/pgf/number format/precision=5},scaled x ticks=false,}} %removing stupid 1e-3 notation

\pgfplotsset{
	wide1/.append style={width=6.5cm, height=3cm,}
}

\pgfplotsset{
	wide2/.append style={width=6.5cm, height=1.5cm,}
}

\pgfplotsset{
	small1/.append style={width=3cm, height=2cm,}
}

\pgfplotsset{default/.style={wide1,
        scale only axis, 
        cycle list/Dark2,
        ylabel near ticks, xlabel near ticks, xmajorgrids, ymajorgrids,
        enlargelimits=false, enlarge y limits=upper, 
        legend pos=north east, 
		every axis legend/.append style={legend cell align=left,align=left, font=\small, fill opacity=0.7,draw opacity=1,text opacity=1},
		every axis plot/.append style={thick, mark=none},
		every axis title/.append style={font=\bfseries},
	}
}

\newcolumntype{L}[1]{>{\raggedright\let\newline\\\arraybackslash\hspace{0pt}}m{#1}}
\newcolumntype{C}[1]{>{\centering\let\newline\\\arraybackslash\hspace{0pt}}m{#1}}
\newcolumntype{R}[1]{>{\raggedleft\let\newline\\\arraybackslash\hspace{0pt}}m{#1}}
\DeclareSymbolFont{sfletters}{OML}{cmbrm}{m}{it}

\DeclareMathSymbol{\salpha}{\mathord}{sfletters}{"0B}
\DeclareMathSymbol{\sbeta}{\mathord}{sfletters}{"0C}
\DeclareMathSymbol{\sgamma}{\mathord}{sfletters}{"0D}
\DeclareMathSymbol{\sdelta}{\mathord}{sfletters}{"0E}
\DeclareMathSymbol{\sepsilon}{\mathord}{sfletters}{"0F}
\DeclareMathSymbol{\szeta}{\mathord}{sfletters}{"10}
\DeclareMathSymbol{\seta}{\mathord}{sfletters}{"11}
\DeclareMathSymbol{\stheta}{\mathord}{sfletters}{"12}
\DeclareMathSymbol{\siota}{\mathord}{sfletters}{"13}
\DeclareMathSymbol{\skappa}{\mathord}{sfletters}{"14}
\DeclareMathSymbol{\slambda}{\mathord}{sfletters}{"15}
\DeclareMathSymbol{\smu}{\mathord}{sfletters}{"16}
\DeclareMathSymbol{\snu}{\mathord}{sfletters}{"17}
\DeclareMathSymbol{\sxi}{\mathord}{sfletters}{"18}
\DeclareMathSymbol{\spi}{\mathord}{sfletters}{"19}
\DeclareMathSymbol{\srho}{\mathord}{sfletters}{"1A}
\DeclareMathSymbol{\ssigma}{\mathord}{sfletters}{"1B}
\DeclareMathSymbol{\stau}{\mathord}{sfletters}{"1C}
\DeclareMathSymbol{\supsilon}{\mathord}{sfletters}{"1D}
\DeclareMathSymbol{\sphi}{\mathord}{sfletters}{"1E}
\DeclareMathSymbol{\schi}{\mathord}{sfletters}{"1F}
\DeclareMathSymbol{\spsi}{\mathord}{sfletters}{"20}
\DeclareMathSymbol{\somega}{\mathord}{sfletters}{"21}
\DeclareMathSymbol{\svarepsilon}{\mathord}{sfletters}{"22}
\DeclareMathSymbol{\svartheta}{\mathord}{sfletters}{"23}
\DeclareMathSymbol{\svarpi}{\mathord}{sfletters}{"24}
\DeclareMathSymbol{\svarrho}{\mathord}{sfletters}{"25}
\DeclareMathSymbol{\svarsigma}{\mathord}{sfletters}{"26}
\DeclareMathSymbol{\svarphi}{\mathord}{sfletters}{"27}

\newcommand{\thickhline}{%
    \noalign {\ifnum 0=`}\fi \hrule height 1pt
    \futurelet \reserved@a \@xhline
}
\newcolumntype{"}{@{\hskip\tabcolsep\vrule width 1pt\hskip\tabcolsep}}
\newtheorem{remark}{Remark}

\title{\LARGE \bf
An Online Stochastic Optimization Approach for Insulin Intensification in Type 2 Diabetes with Attention to Pseudo-Hypoglycemia*\\

\thanks{*This work was funded by the IFD Grand Solution project ADAPT-
	T2D, project number 9068-00056B.}
}
\author{Mohamad Al Ahdab$^{1}$, Torben Knudsen$^{1}$, Jakob Stoustrup$^{1}$, John Leth$^{1}$% <-this % stops a space
	\thanks{$^{1}$The authors are with Section of Automation and Control, Department of Electronic Systems, Aalborg University, Aalborg Øst, Denmark
		{\tt\small \{maah,tk,jakob,jjl\}@es.aau.dk}}%%%
}

\begin{document}

\maketitle
\thispagestyle{empty}
\pagestyle{empty}

%%%%%%%%%%%%%%%%%%%%%%%%%%%%%%%%%%%%%%%%%%%%%%%%%%%%%%%%%%%%%%%%%%%%%%%%%%%%%%%%
\begin{abstract}
In this paper, we present a model free approach to calculate long-acting insulin doses for Type 2 Diabetic (T2D) subjects in order to bring their blood glucose (BG) concentration to be within a safe range. The proposed strategy tunes the parameters of a proposed control law by using a zeroth-order online stochastic optimization approach for a defined cost function. The strategy uses gradient estimates obtained by a Recursive Least Square (RLS) scheme in an adaptive moment estimation based approach named AdaBelief. Additionally, we show how the proposed strategy with a feedback rating measurement can accommodate for a phenomena known as relative hypoglycemia or pseudo-hypoglycemia (PHG) in which subjects experience hypoglycemia symptoms depending on how quick their BG concentration is lowered. The performance of the insulin calculation strategy is demonstrated and compared with current insulin calculation strategies using simulations with three different models. 
% using a feedback rating measurement obtained from the subjects. for Upon the initiation of insulin treatment for subjects with type 2 diabetes (T2D), symptoms of hypoglycemia can be experienced by the subjects depending on how quick their blood glucose concentrations are reduced. This phenomena is known as relative hypoglycemia or pseudo-hypoglycemia (PHG) and it is different from patient to patient. In this paper, we propose a model free strategy to calculate insulin doses for T2D which mitigates PHG. The strategy makes use of a feedback rating mechanism obtained from the subjects. The proposed strategy tunes the parameters of a proposed control law by using an online stochastic optimization approach for a defined cost function. The strategy uses gradient estimates obtained by a Recursive Least Square (RLS) scheme in an adaptive moment estimation based approach named AdaBelief. The performance of the insulin calculation strategy is demonstrated and compared with current insulin calculation strategies using simulations with three different models.
\end{abstract}
\section{Introduction}
Subjects with type 2 diabetes (T2D) experience elevated levels of blood glucose (BG) concentrations known as hyperglycemia due to an imbalance between their insulin secretion rate and the effectiveness of insulin to lower glucose concentration. If high BG concentrations are left untreated, subjects can develop complications such as cardiovascular diseases, eyesight damage, and more. The treatment procedure for T2D initially begins with lifestyle changes and oral medications. However, when these methods are insufficient to lower BG concentrations, T2D subjects can begin to administer long acting insulin, for example once daily using insulin pens, based on self monitored blood glucose measurements (SMGB) of Fasting BG (FBG). The insulin treatment initially aims at finding the optimal insulin (insulin intensification/titration) dose to keep BG concentration within a safe range. This process is clinically challenging since subjects with T2D are different on a behavioural and a physical level. Moreover, administrating too much insulin can lead to low BG levels known as hypoglycemia which can cause blurred vision, fainting, or death in severe cases. On the other hand, not administrating enough insulin will cause the subject to remain in hyperglycemia for extensive periods of time. In addition to these challenges, T2D subjects can experience symptoms of hypoglycemia even when they have a BG level above the clinical level of hypoglycemia. This phenomena is referred to as relative hypoglycemia or Pseudo-HypoGlycemia of Type I (PHG) \cite{case2019pseudopheochromocytoma,seaquist2013hypoglycemia}. PHG happens when T2D patients reduce their FBG aggressively after staying at a fixed level for a period of time. Due to these challenges, several attempts were made to use automated insulin dose calculators for T2D. Standard of care insulin guidance algorithms such as the ones in \cite{kadowaki2017insulin} are based on SMBG measurement to decide on a fixed insulin dose weekly. These titration strategies can take a long time to bring FBG concentrations to a safe level. While this can be beneficial to avoid PHG, it is still conservative since T2D subjects are different from each other and long titration periods can be limiting for subjects which can have their FBG levels lowered more quickly. Other titration algorithms based on control theory exists in the literature such as \cite{aradottir2019model} which is model based and \cite{krishnamoorthy2020model} which is model free. The work in \cite{aradottir2019model} relies on a model which can be limiting and challenging to apply for a wide range of T2D subjects. Additionally, the algorithms lower the FBG concentration aggressively which can be problematic for PHG. On the other hand, the work in \cite{krishnamoorthy2020model} proposed to use an Extremum Seeking Control (ESC) strategy to alleviate the need for a detailed model of T2D subjects and demonstrated the effectiveness of such approach. Nevertheless, the strategy was tested against one model only and with limited variation on the parameters without measurement noise. Additionally, the strategy lowers the FBG aggressively for all subjects without consideration for PHG. The contributions in this paper are as follows:
\begin{itemize}
 \item We propose a model free strategy which handles measurement noise on SMBG. Additionally, we test the strategy for three different models. Namely, the model which was used in \cite{krishnamoorthy2020model}, an extended version of it from \cite{ccta}, and a model based on the high fidelity model \cite{al2021glucose}. The strategy was shown in simulation to be more robust to parameter variations than the recently proposed model free approach in \cite{krishnamoorthy2020model}.
 \item We investigate the possibility of designing our strategy to handle PHG in insulin titration, which to our knowledge, never has been done before. The idea for handling PHG is inspired by the recent works of including human ratings as feedback in control strategies as done in \cite{humanfeedback}. We propose to use a score in the calculation of insulin doses, provided by the T2D subjects, and/or their medical professionals on a daily basis reflecting their well-being with respect to PHG symptoms.
 \item We propose a zeroth-order online optimization approach for a defined cost to tune the parameters of a chosen feedback control law. The method uses the recently proposed adaptive moment estimation algorithm AdaBelief \cite{zhuang2020adabelief} with gradient information provided by a RLS.
\end{itemize}
The paper is structured as follows. Section \ref{sec:ProblemSpecs} explains the setup of the problem. Sections \ref{sec:RLS} and \ref{sec:GDS} provide a description on a directional forgetting RLS and the AdaBelief strategy in the context of tuning the control law parameters, respectively. Section \ref{sec:CostFuncdef} then defines the cost functions which we aim to minimize in order to tune the control law parameters. After that, we propose a simulation model for PHG in section \ref{sec:PHGmodel} and provide a discussion on the used glucose-insulin models for simulation in section \ref{sec:simmodels}. Finally, we present the simulation results in section \ref{sec:Res} and provide a conclusion in section \ref{sec:conc}.
\section{notations}
The symbol $:=$ indicates "defined by". All vectors are considered as column vectors, $\|\cdot\|$ denotes the 2-norm, and $~^\mathrm{T}$ denotes transpose. All probabilistic considerations in this paper will be with respect to an underlying probability space $(\Omega,\mathcal{F},\mathbb{P})$ and \textit{every statement will be understood to be valid with probability 1}. We let $L^2_l=L^2_l(\Omega,\mathcal{F},\mathbb{P})$ denote the set of $l$-valued measurable maps $f:\Omega\to\mathbb{R}^l$ with $\mathbb{E}[\|f\|^2]<\infty$. For a random variable $x$ we write $\mathsf{x}=x(\omega)$ for the realization of the random variable. For probability distributions, we use $Beta(\alpha,\beta)$ to denote the beta distribution with parameters $\alpha$ and $\beta$, $\mathcal{N}(\mu,\Sigma)$ to denote the normal distribution with mean $\mu$ and covariance $\Sigma$, $\mathcal{U}(a,b)$ for a continuous uniform distribution with bounds $a$ and $b$, and $\mathcal{U}\{a,b\}$ for a discrete uniform distribution with bounds $a$ and $b$.
%For a random variable $x$, the value of it will be denoted as $\mathrm{x}$ such that $x=\mathrm{x}\sim p_{x}(\mathrm{x})$ with $p_{x}$ being the probability density function for $x$. 
If the difference between two consecutive time instants $t_{k}$ and $t_{k+j}$ is such that $t_{k+j}-t_{k}=jT,~j,k\in\mathbb{N}$ with $T\in\mathbb{R}$ being a constant, then variables that are indexed with time $x(t_{k}),x(t_{k+j})$ will be denoted by $x(k),x(k+j)$ for ease of notation. %Additionally, if we have $t_{k+j}-t_{k}=jT_{1},~j,k\in\mathbb{N}$ and $t_{m+j}-t_{m}=jT_{2},~j,m\in\mathbb{N}$ such that $T_{2}<T_{1}$, then we denote $x(t_{m_{k}})=x(m_{k})$ the value $x(m)$ which is closest in time to $x(k)$, i.e. $t_{m_{k}}=\mathrm{argmin}_{t_{m}}|t_{m}-t_{k}|$. 
We write $\{a:s:b\}$ for a sequence of numbers going from $a$ to $b$ equally spaced by $s$. 
%For a collection of $n\in\mathbb{N}$ variables $\{x(i)\}^{n}_{i=1}$ or $\{x(t_{i})\}^{n}_{i=1}$,  the notation $x^{1:n}$ is used for both of them. Moreover, 
%If $x(i)\in\mathbb{R}^m$ for each element of $\{x(i)\}^{n}_{i=0}$, then $x^{0:n}:=\left[x(0)^T,\dots,x(n)^T\right]^{\mathrm{T}}\in\mathbb{R}^{m(n+1)}$.
We let $[a,b]$ denote the closed interval from $a$ to $b$, and $[a~b]$ denote the row vector with coordinates $a$ and $b$. For a diagonal matrix $A$ with diagonal entries $a=[a_{1}\cdots a_{n}]^\mathrm{T}$, the notation $A=\text{diag}(a)$ is used. The symbol $\mathrm{I}_{n}$ is used to denote the $n\times n$ identity matrix and the symbol $\bf{1}$ is used to denote a vector of ones. Finally, a projection operator is defined as $\Pi_{\Theta,\Sigma}(x):=\text{argmin}_{\theta\in\Theta}\|\Sigma^{1/2}(\theta-x)\|$ with $\Sigma$ a positive definite matrix and $\Theta$ a compact set.
\section{Control Strategy}
\subsection{Problem Specification}
\label{sec:ProblemSpecs}
In this section we present the aim and the proposed strategy for insulin titration. We assume that the insulin-glucose dynamics of a T2D subject can be modeled according to the following general form 
\begin{subequations}\label{dyn}
\begin{align}
    \label{eq:Dynamics}
    x(k+1) &= f(x(k),\Delta u(k),w(k)),\\
    \label{eq:ControlLaw}
    \Delta u(k) &=\frac{K_{p}}{1+K_{s}e_{s}(k)}e_{g}(k),\\
    \label{eq:Öutput}
    y(k) &= h(x(k),v(k)),\\
    \label{eq:costfunction}
    z(k) &= c(y(k)),
\end{align}
\end{subequations}
where $x(k)\in\mathbb{R}^n$ are internal states, $w(k)$ being a sufficiently regular stochastic process (see Remark~\ref{remwv}),
$\Delta u(k)$ is the change of the insulin dose size $u(k)~\si{[U]}$ at day $k$ such that $u(k)=\max \left( \Delta u(k)+u(k-1),0\right)$ with the feedback control law \eqref{eq:ControlLaw} parameterized with $\theta=[K_{p}~K_{s}]^{\mathrm{T}}\in\Theta\subseteq\mathbb{R}^{2}$, $y(k)=[y_{g}(k) ~y_{s}(k)]^\mathrm{T} \in \mathbb{R}^2$ represents the SMBG measurement $y_{g}(k)$ and the PHG score $y_{s}(k)$ at day $k$, the measurement noise $v(k)$ is an i.i.d. stochastic process independent of $w(k)$, $e_{g}(k):=r-y_{g}(k)$ with $r$ being a reference, $e_{s}(k):=(H-y_{s}(k))/H$ with $H\in\mathbb{R}_{>0}$ being the maximum score for a PHG scale used by the subjects as a feedback method for their hypoglycemia symptoms. The maximum score $H$ means no hypoglycemia symptoms were experienced by the subjects. See Section \ref{sec:PHGmodel} for more details. The variable $z(k)\in\mathbb{R}$ is the value of a cost function $c(y(k))$ we desire to minimize. We write $x(k;\theta)$, $y(k;\theta)$ and $z(k;\theta)$ whenever the dependency on the control parameter is relevant. 
\begin{remark}\label{remwv}
The structure in \eqref{dyn} represents a vast variety of models in the current literature e.g., the ones in \cite{krishnamoorthy2020model,ccta} where $w$ represents white noise in \cite{krishnamoorthy2020model}, a jump process in \cite{ccta}, and $v$ represents white noise in \cite{ccta}.  
\end{remark}
We assume that the functions $f,h,g$ are sufficiently regular e.g., Lipschitz continuous which is a typical assumptions for biological systems. Now for ease of notation let $q$ denote either $x$ or $y$. We then assume that there exists\footnote{In application/simulation $\Theta$ can often be obtained by a conservative guess.} $\Theta\subseteq\mathbb{R}^{2}$ with $0\in\Theta$ such that for every $\theta\in\Theta$ we have $q(k;\theta)\in L^2_l,~(l=2,n)$, $\|q(k;\theta)\|\leq\tilde{q}$ for some $\tilde{q}\in L^2_1$, and 
$\lim_{k\to\infty}q(k;\theta)=q^*$ with probability 1. Note that $q^*$ depends on $\theta$, and that by dominated convergence we obtain $q^*\in L^2_l$, $\lim_{k\to\infty}\mathbb{E}[\|q(k)-q^*\|^2]=0$, and $\lim_{k\to\infty}\mathbb{E}[q(k)]=\mathbb{E}[q^*]$. 
%Note that $q(k)$ depends on $\theta(k-1)$ and we will write $q(k,\theta(k-1))$ for clarity when it is relevant. 
Let $\bar{c}(k,\theta):=c\left(y\left(k;\theta\right)\right)+c_{\theta}(k,\theta)$ with $c_{\theta}(k,\theta)$ being a known (in closed form) differentiable cost in $\theta$,
we aim to find a sequence of estimates $\{\hat{\theta}(k)\}_{k\in \mathbb{N}}$ in $\Theta$ which tracks the sequence $\{\theta^{*}(k)\}_{k\in \mathbb{N}}$ that solves the following 
\begin{equation}
    \label{eq:MainProb}
    \theta^{*}(k) = \underset{\theta \in \Theta}{\operatorname{argmin}}~\bar{c}(k+1,\theta).
\end{equation}
This problem can be thought of as a tracking problem where a pursuer $\hat{\theta}(k)$ tries to track a target $\theta^{*}(k)$. For each $\hat{\theta}(k)$, an inexact gradient $\hat{g}(k+1)$ will be estimated based on $\bar{c}\left(k+1,\hat{\theta}(k)\right)$. The pursuer will then use $\hat{g}(k+1)$ to obtain an estimate $\hat{\theta}(k+1)$. This problem is known as zeroth-order online optimization in the bandit setting. The term zeroth-order refers to the fact that for every estimate $\hat{\theta}(k)$ we only obtain a cost function value information. Note that by assumption, the sequence $\theta^{*}(k)$ will converge (with probability 1) to a random variable $\theta^{*}$.
% Note that at $q^*$, we have $\theta^{*}(k)=\theta^{*}$ becomes constant with $\nabla_\theta\bar{c}(k,\theta^{*})=0$. Meaning that the sequence $\{\theta^{*}(k)\}$ will converge to a constant $\theta^{*}$. 
See the works in \cite{flaxman2004online,onlinestoch, bedi2018tracking} for convergence analysis in a related setting.
%The work in \cite{onlinestoch} derived convergence bounds for this problem with an online stochastic gradient decent strategy under the assumption that the the hessian $\nabla_{\theta}^{2}\bar{c}\left(k,\theta\right)$ is bounded for all $k$, and that the gradient estimate $\hat{g}(k)$ is a sufficiently well estimate of $\nabla_{\theta}\bar{c}(k,\theta(k)))$ ( $\mathbb{E}\left[e(k)\right]=0$ and bounding conditions on $\mathbb{E}\left[\norm{e(k)}^2\right]$, see also \cite{bedi2018tracking} for more details).
For the work in this paper, we use an adaptive moment based method named AdaBelief \cite{zhuang2020adabelief} for a gradient based optimization as detailed in section \ref{sec:GDS}. For the gradient estimates, we assume a local linear model for the cost $z(k)=c\left(y(k,\hat{\theta}(k-1))\right)$.
\begin{equation}
    \label{eq:FirstOrder_approx}
    z(k;\hat{\theta}(k-1)) \approx [\hat{\theta}^{\mathrm{T}}(k-1)~\boldsymbol{1}^{\mathrm{T}}]\begin{bmatrix}g_{z}(k)\\b(k)\end{bmatrix}:=\phi^{\mathrm{T}}(k)\psi(k),
\end{equation}
where $g(k)=g_{z}(k)+\nabla_{\theta}c_{\theta}(k,\hat{\theta}(k-1))$ represents an approximate for the gradient $\nabla_{\theta}\bar{c}(k,\hat{\theta}(k-1))$, and  $b(k)$ is a bias term. A recursive least squares (RLS) strategy can then be used to obtain an estimate $\hat{g}_{z}(k)$ as described in section \ref{sec:RLS}.

%If the gradient decent scheme in section \ref{sec:GDS} updates the parameters $\theta$ slowly enough such that the change of insulin doses between the subsequent days is limited or constrained (e.g. saturating the change such that $|\Delta u(k)|\leq 8~\si{[U]}$), the cost $z(k)$ will evolve slowly enough such that subsequent parameters $\hat{hat}$ will be close to each other and thus the problem approximates

%the process $z(k)$ will be close enough to $c_{str}(\theta(k))$ (in the mean sense) such that the gradient estimate $g(k)$ is a good approximation to $\nabla_{\theta}\bar{c}_{st}(\theta(k))$. This is a well-known discussed result in the ESC literature, e.g. see \cite{tan2010extremum}. To ensure this, we also use $|\Delta u(k)|\leq 8~\si{[U]}$ to limit the rate of change for insulin. The limitation of the rate of change is also desired since it reduces the complexity for the dosing scheme which helps with subjects adhering to the treatment \cite{sarbacker2016adherence}.

\subsection{Estimating the gradient with RLS}
\label{sec:RLS}
For the estimation of the gradient, \eqref{eq:FirstOrder_approx} is used in an RLS with exponential forgetting setting. Least square estimation with exponential forgetting aims at finding the value $\hat{\spsi}$ which minimizes \hspace{-0.5mm}{\small$\sum_{i=0}^{k} \lambda^{k-i}\left(\mathrm{z}(i)-\sphi^{\mathrm{T}}(i)\hat{\spsi}(i)\right)^{\mathrm{T}}\left(\mathrm{z}(i)-\sphi^{\mathrm{T}}(i)\hat{\spsi}(i)\right)$} with $\lambda\in(0,1]$ being a forgetting factor. The forgetting factor is used to put more emphasis on recent incoming data when compared to old one. This makes it useful for estimating time varying parameters such as the gradient $g(k)$ which we aim to estimate. Additionally, it is known that without persistent excitation in $\phi(k)$ (see \cite{goel2020recursive} for more details), the covariance $P:=\mathbb{E}\left[(\psi-\hat{\spsi})^{\mathrm{T}}(\psi-\hat{\spsi})\right]$ can become unbounded. We can ensure persistence by adding a small dither to our control law parameters $\hat{\theta}(k)$. Moreover, to further ensure the boundedness of the covariance matrix regardless of the persistent excitation condition, we apply the directional forgetting RLS algorithm proposed in \cite{cao2000directional}.
%The directional forgetting RLS decomposes the information matrix $R(k-1):=P^{-1}(k-1)$ into $R(k-1)=R_{1}(k-1)+R_{2}(k-1)$ where $R_{1}(k-1)$ is chosen such that $R_{1}(k-1)\sphi(k)=0$ and $R_{1}(k-1)\neq 0$. The forgetting in the strategy is then applied only to $R_{2}(k-1)$. %The directional forgetting RLS decomposes the information matrix $R(k):=P^{-1}(k)$ into $R(k)=R_{1}(k)+R_{2}(k)$ such that $R_{1}(k)\sphi(k)=0$ when $\sphi(k)\neq0$, and applies the forgetting on the "information rich direction" $R_{2}(k)$.
%which performs forgetting only on the information rich directions coming by $\phi(k)$.
The recursive estimation is summarized in Algorithm~\ref{alg:RLS}. Note that the algorithm includes an update step for the information matrix $R(k)$ separately from $P(k)$ to avoid computing $P^{-1}(k)$ in the calculation of $\bar{P}(k)$ and $M(k)$.
{\vspace{0.05mm}
\begin{algorithm}
\label{alg:RLS}
\DontPrintSemicolon
  \KwInput{Estimates $\hat{\spsi}(k-1)$ with covariance matrix $P(k-1)$ and $R(k-1)=P^{-1}(k-1)$, regressor $\sphi(k)$ and measurement $\mathrm{z}(k)$, forgetting factor $\lambda\in(0,1]$, and a threshold $\epsilon_{\phi}$ (chosen to be in the order of the minimum added dither) to stop forgetting when $\|\sphi(k)\|<\epsilon_{\phi}$.}
  \KwOutput{$\hat{g}(k)$ as a component in $\hat{\spsi}(k)$, $P(k)$ and $R(k)$.}
  \uIf{$\|\sphi(k)\|\geq\epsilon_{\phi}$}{
  $\bar{P}({k})=P({k-1})+\frac{1-\lambda}{\lambda}\left(\sphi({k})R(k-1) \sphi^{\mathrm{T}}({k})\right)^{-1} \sphi^{\mathrm{T}}({k}) \sphi({k})$\;
  $M(k)=(1-\lambda) \frac{R(k-1) \sphi(k) \sphi^{\mathrm{T}}(k)}{\sphi^{\mathrm{T}}(k) R(k-1) \sphi(k)}$\;}
  \Else {$\bar{P}(k) = P({k-1})$\;
  $M(k)=\mathrm{0}$\;}
  $K_{f}(k)=\bar{P}(k) \sphi(k)\left(1+\sphi^{\mathrm{T}}({k}) \bar{P}({k}) \sphi({k})\right)^{-1}$\\
  $\hat{\spsi}(k) = \hat{\spsi}(k-1) + K_{f}(k)\left(\mathrm{z}(k)-\sphi^{\mathrm{T}}(k)\hat{\spsi}(k-1)\right)$\\
  $P({k})=\bar{P}({k})-\bar{P}({k}) \sphi({k})\left(1+\sphi^{\mathrm{T}}({k}) \bar{P}({k}) \sphi({k})\right)^{-1} \sphi^{\mathrm{T}}({k}) \bar{P}({k})$\;
  $R(k) = (\mathrm{I}-M(k))R(k-1)+\sphi(k)\sphi^{\mathrm{T}}(k)$
\caption{RLS with directional forgetting}
\end{algorithm}}

The matrix $M$ in the RLS strategy applies the forgetting factor $\lambda$ on a subspace of the column space of the information matrix $R$, for details see \cite{cao2000directional}. 
\subsection{Gradient Decent Strategy}
\label{sec:GDS}
Due to the stochastic nature of the problem and the fact that our gradient estimates are noisy, we propose to use a stochastic optimization method with an adaptive step size. Adaptive moment based strategies such as Adam and its variants \cite{alacaoglu2020new} have gained wide interest in the field of deep learning as methods to perform stochastic optimization. Additionally, the work in \cite{wu2020flight} proposed to use the original Adam in an ESC scheme to adapt the step size based on the estimated gradient. However, the original Adam can diverge even for a convex optimization problem \cite{alacaoglu2020new}.
In this work, we propose to use AdaBelief, a variant of Adam \cite{zhuang2020adabelief}. In \cite{zhuang2020adabelief}, AdaBelief was shown to combine the fast convergence of Adam based strategies with the good generalization of stochastic gradient decent strategies.
The online stochastic optimization based AdaBelief strategy (AdaOS) is presented in Algorithm \ref{alg:ESCAdamh}.
{\begin{algorithm}
\label{alg:ESCAdamh}
\DontPrintSemicolon
\Parameter{
    Parameter $\alpha>0$, smoothing parameters $0\leq\beta_{1}\leq1$ and $0\leq\beta_{2}\leq1$, vector of small numbers $\epsilon$, and projection $\Pi_{\Theta,\Sigma}(x)=\text{argmin}_{\theta}\|\Sigma^{1/2}(\theta-x)\|$.
    Note that all the operations in the algorithm are element-wise.}
  \KwInput{initial moments $m(0)=0$ and $s(0)=0$.} 
  \KwOutput{$\hat{\theta}(k)$}
  k=0\;
  \While{Ongoing Titration}{
  $k \leftarrow k+1$\;
  Run RLS to obtain $\hat{\mathrm{g}}_{z}(k)$.\;
  $\hat{\mathrm{g}}(k)=\hat{\mathrm{g}}_{z}(k)+\nabla_{\theta}c_{\theta}(\hat{\theta}(k-1))$.\;
  $m(k) = \beta_{1}m(k-1)+(1-\beta_{1})\hat{\mathrm{g}}(k)$\;
  $s(k) = \beta_{2}s(k-1) + (1-\beta_{2})(m(k)-\hat{\mathrm{g}}(k))^2+\epsilon$,\;
  $\hat{m}(k) = \frac{m(k)}{1-\beta^{k}_{1}}, \hat{s}(k) = \frac{s(k)}{1-\beta^{k}_{2}},~\text{(Bias-Correction.)}$\;
  $\hat{\theta}(k) = \Pi_{\Theta,\text{diag}\left(\hat{s}(k)\right)}\left(\hat{\theta}(k-1) - \alpha\frac{\hat{m}(k)}{\sqrt{\hat{s}(k)}+\epsilon}\right)$ }
\caption{AdaOS algorithm}
\end{algorithm}
}
To get an intuition of how AdaOS works, we note that $m(k)$ is an exponential moving average (the output of a first order low pass filter) for the gradient estimate $\hat{\mathrm{g}}(k)$. Thus, the algorithm produces a smoother version $\hat{m}(k)$ of the estimated gradient $\hat{g}(k)$. As for $s(k)$, it reflects the difference between the gradient estimate $\hat{\mathrm{g}}(k)$ and our "belief" $m(k)$ such that, for an increased difference, the stepping size $\alpha/\left(\sqrt{\hat{s}(k)}+\epsilon\right)$ will decrease and vice versa. In this paper, the parameters for the algorithm \label{alg:ESCAdam} are chosen to be $\alpha=10^{-3},~\beta_{1}=0.99,~\beta_{2}=0.999$ and $\epsilon=10^{-8}\bf{1}$ which are the typical parameters used in \cite{zhuang2020adabelief} for AdaBelief and Adam based strategies in practice \cite{alacaoglu2020new}. Additionally, we choose $\Theta=[0~2]^2$ for the control law parameters.\footnote{Simulation results show that all parameters in $\Theta$ give rise to a stable behaviour.} The step in line 7 of the algorithm is used to correct for the initialization bias.
\subsection{Cost function definition}
\label{sec:CostFuncdef}
The main aim of the control strategy is to bring the glucose concentration $y_g(k)$ to a safe level. For this objective, we propose the following cost function
\begin{equation}
    \label{eq:refpenality}
    c_{g}(k) = \left(e_{g}(k)/r\right)^2,
\end{equation}
Note that the division by $r$ was made to scale $c_{g}(k)$ to be of order 1. The safe range of FBG is chosen to be between $4~\si{[mmol/L]}$ and $6~\si{[mmol/L]}$ according to the standard of care for insulin titration strategies \cite{kadowaki2017insulin}. Therefore, we choose the reference $r=5.5~\si{[mmol/L]}$. Note that the reference is chosen to be larger than the middle of the range $[4~6]~\si{[mmol/L]}$ since hypoglycemia (FBG concentrations below $4~\si{[mmol/L]}$) are more dangerous than hyperglycemia.
Additionally, we use the following cost to penalize FBG concentrations which are within the hypoglycemic range
\begin{equation}
    \label{eq:hyppenality}
    c_{h}(k) = \text{softmin}\left(e_{g}(k),0\right)^2,
\end{equation}
where $\text{softmin}$ is the soft minimum function.\footnote{$\text{softmin}(x_{1},x_{2}) = -\frac{1}{a}\log(\exp(-ax_{1})+\exp(-ax_{2}))$, with $a$ being a constant chosen as 50 in this paper.}
%\begin{equation}
%    \label{eq:hyppenality}
%    c_{h}(k) = \exp(-(y_{g}(k)-5))-1.
%\end{equation}
Moreover, to keep the PHG score $y_{s}$ as high as possible, the following is used
\begin{equation}
    \label{eq:scorepenality}
    c_{s}(k) = \left(e_{s}(k)\right)^2.
\end{equation}
The cost in measurements is then chosen as $c(y(k)) = c_{g}(k) + 10c_{h}(k) + 10c_{s}(k)$. In addition to the cost in measurement, we include a cost which is more related to our setup of the optimization scheme. Namely, we consider the cost $c_{\theta}(k,\theta)=0.5\left\|\theta - \hat{\theta}(k-2)\right\|^2$ in order to ensure a smooth change in the decision variables between iterations and to ensure that $\hat{\theta}(k)=\hat{\theta}(k-1)$ when $e_{s}(k)=0$ and $e_{g}(k)=0$. Finally, the total cost is $\bar{c}(k,\theta)=c\left(y(k)\right)+c_{\theta}(k,\theta)$. 
%Note that the gradient of the added term is $\left(\theta-\hat{\theta}(k-2)\right)$ and can be summed directly to the estimated gradient of $c(y(k))$ in algorithm \ref{alg:ESCAdamh}.
%\begin{equation}
 %   \label{eq:TotalCost}
%    c(y(k)) = %\frac{1}{k}\sum^{k}_{i=0}
%    c_{g}(k) + c_{h}(k) + c_{s}(k).
%\end{equation}

\section{Simulation Models}
In this paper, we use simulations in order to test and validate the developed titration strategy.
In this section, we first describe the development of a model to simulate the PHG scores provided by T2D subjects during their treatment in Section \ref{sec:PHGmodel}. Afterwards, we describe three different models used to simulate the glucose-insulin dynamics in Section \ref{sec:simmodels}.
\subsection{PHG score model}
\label{sec:PHGmodel}
\begin{figure}
    \centering
    \includegraphics[width=0.5\textwidth]{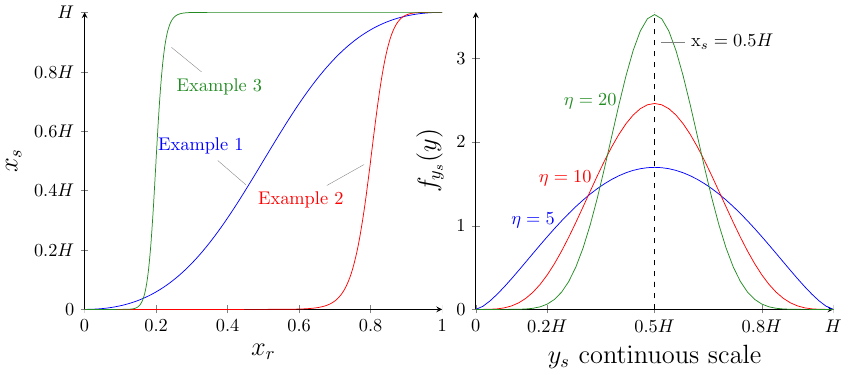}
    \caption{\textbf{Left}: Three different examples of the noise-free PHG score $x_{s}$. {\color{blue} Example 1}: $\rho=2$ and $d=0.5$. {\color{red} Example 2}: $\rho=5$ and $d=0.8$. {\color{ForestGreen} Example 3}: $\rho=20$ and $d=0.2$. \textbf{Right}: The density function of $y_{s}$ with a continuous scale given $\mathrm{x}_{s}=0.5H$ for different values of $\eta$.}
    \label{fig:PHGexamples}
\end{figure}

For the PHG scores, we assume that at each day $k$ the T2D subjects will provide a score $y_{s}(k)\in[0,H]$ if a continuous scale is used or $y_{s}(k)\in\{0,\dots,H\}$ if a discrete scale is used. If the subjects were experiencing no hypoglycemia symptoms then they would provide the maximum score $H$. On the other hand, if the subjects were experiencing severe hypoglycemia symptoms then they would provide the minimum score $0$. The determination of the range of symptoms and what they correspond to on the scale can be assigned by the medical professionals. See the study in \cite{divilly2022hypo} for an example.
We intend in this section to develop a general simulation model for PHG scores which can then be used together with simulated T2D subjects. This model is used to test if the strategy can work with a feedback score which correlates with how rapid the BG concentration is lowered. The PHG simulation model should also take into account that subjects may react differently to how rapid their BG is being lowered.
Following the observations that patients who have been staying at a high BG concentration level for a period of time can develop hypoglycemia like symptoms when BG are decreased aggressively, we first define the BG decrease ratio $x_{r}(k)$ as following
%\begin{subequations}
    \begin{align}
    x_{r}(m):=\text{max}\left(\frac{x_{g}(m)}{\mu(m)},1\right),~~
    \mu(m) = \frac{1}{h^{'}}\sum_{i=m-h^{'}}^{m}x_{g}(i),
    \end{align}
%\end{subequations}
where $x_{g}(m)~\si{[mmol/L]}$ is the BG concentration at minute $mT_{m}$ with $T_{m}$ being a sampling time in the order of minutes, and $h^{'}=\frac{24\times60}{T_{m}}h$ with $h~[\si{Day}]$ being a time window for the moving average $\mu(m)$ which captures the history of the BG concentration for the T2D subjects. If the BG levels do not change significantly when compared to the moving average $\mu(m)$ then the value of $x_{r}(m)$ is close to 1. However, when the BG level drops significantly compared to the previous history of BG levels (captured in the moving average $\mu(m)$), the value of $x_{r}(m)$ will be closer to 0. The BG decrease ratio $x_{r}(m)$ models the aggressiveness of lowering BG concentration. Now, in order to also take into account that subjects with T2D react differently to the drop of their BG concentration, we define the function $\text{sig}_{\rho,d}: [0,1]\rightarrow [0,1];~x\mapsto \text{sig}_{\rho,d}(x)$ as
\begin{equation}
    \label{eq:PHGscoresig}
    \text{sig}_{\rho,d}(x):=\begin{cases}\frac{1}{1+\left(\frac{x^{-\log(2)/\log(d)}}{1-x^{-\log(2)/\log(d)}}\right)^{-\rho}},~x\in[0,1)\\
    1, ~ x=1\end{cases}
    %\text{sig}_{H,\rho,c}(x):=H\left(1+\left(\frac{x^d}{1-x^d}\right)^{-\rho}\right)^{-1},
\end{equation}
with $\rho$ being a constant representing the sensitivity for different $x_{r}(k)$, and $d$ is the value such that $\text{sig}_{\rho,d}(d)=0.5$. Finally, the noise-free PHG score is defined as
\begin{equation}
    \label{eq:PHGscore_day_k}
    x_{s}(k):= H\text{sig}_{\rho,d}(x_{r}(k))
\end{equation}
Figure \ref{fig:PHGexamples} shows three different examples of noise-free $x_{s}$ versus BG decrease ratios $x_{r}$ for three different subjects. In Example 1, the range of BG decrease ratio $x_{r}$ in which the subject reacts to with different scores is the widest ($\rho=2$), While in Example 3, the subject has the narrowest range ($\rho=20$). In Example 2, the subject has the lowest tolerance for BG decrease ratio ($d=0.8$), while the subject in Example 3 has the highest tolerance ($d=0.2$). With the shape parameters $d$ and $\rho$, one can construct a wide variety of sigmoidal curves which enables us to model different possibilities of subjects reacting to their BG decrease ratio.  
To model noises and disturbances on the PHG score, let $\zeta(k)\sim Beta \left((x_{s}(k)/H)\eta,(1-x_{s}(k)/H)\eta\right)$, then the PHG score measurements are $y_{s}(k)=H\zeta(k)$ if continuous scales are used, or $y_{s}(k)=\text{round}\left(H\zeta(k)\right)$ if discrete scales are used. Note that given the realization $x_{s}(k) = \mathrm{x}_s$, then $\mathbb{E}[H\zeta(k)]=\mathrm{x}_s$ and $\text{Var}(H\zeta(k))=\frac{\mathrm{x}_s(H-\mathrm{x}_s)}{1+\eta}$, this means that the parameter $\eta$ can be viewed as a precision parameter in the sense that for a fixed $x_{s}(k)$, the larger $\eta$ is the smaller is the variance and vice versa. Figure \ref{fig:PHGexamples} shows the probability density function of a continuous scale $y_{s}$ given $\mathrm{x}_{s}=0.5H$. 
% \begin{figure}
%     \centering
%     \begin{tikzpicture}[scale=0.5]
%         % define macros which are needed for the axis limits as well as for
%         % setting the domain of calculation
%         \pgfmathsetmacro{\xmin}{0}
%         \pgfmathsetmacro{\xmax}{1}
%     \begin{axis}[
%     axis lines = left,
%     xlabel = {$y_{s}$ continuous scale},
%     ylabel = {$f_{y_{s}}(y)$},
%     xmin=0, xmax=1,
%     no markers,
%     xtick={0,0.2,0.5,0.8,1},
%     xticklabels={$0$,$0.2H$,$0.5H$,$0.8H$,$H$},
%     ]
%         \addplot gnuplot [raw gnuplot,blue] {
%             \GnuplotDefs
%             plot [x=0:1] beta(x,0.5*5,(1-0.5)*5);
%         };
%         \node [align = center] at (axis description cs:0.16,0.3) {\color{blue}{$\eta=5$}};
%         \addplot gnuplot [raw gnuplot,red] {
%             \GnuplotDefs
%             plot [x=0:1] beta(x,0.5*10,(1-0.5)*10);
%         };
%         \node [align = center] at (axis description cs:0.25,0.44) {\color{red}{$\eta=10$}};
%         \addplot gnuplot [raw gnuplot,ForestGreen] {
%             \GnuplotDefs
%             plot [x=0:1] beta(x,0.5*20,(1-0.5)*20);
%         };
%         \node [align = center] at (axis description cs:0.32,0.7) {\color{ForestGreen}{$\eta=20$}};
%         \addplot[thick, samples=50, smooth,domain=0:1,dashed] coordinates {(0.5,0)(0.5,3.55)};
%         \node [pin=0:{$\mathrm{x}_{s}=0.5H$}] at (axis description cs:0.5,0.9) {};
        
%     \end{axis}
% \end{tikzpicture}
%     \caption{The density function of $y_{s}$ with a continuous scale given $\mathrm{x}_{s}=0.5H$ for different values of $\eta$.}
%     \label{fig:distys}
% \end{figure}
Additionally in simulation, if T2D subjects report a lower score when their BG is actually in the hypoglycemia region, then the PHG score is ignored since it is clearly not a case of PHG. 
\subsection{Glucose-Insulin Simulation Models}
\label{sec:simmodels}
For the glucose-insulin dynamic simulations in this paper, we consider three different simulation models. The first model, denoted  "Model 1", is the same model used in \cite{krishnamoorthy2020model}. Model 1 considers FBG only and it will be used in Section \ref{sec:CompESC} for a detailed comparison with the insulin titration strategy presented in \cite{krishnamoorthy2020model}. As for the second model, denoted "Model 2", we use an extension of Model 1 in order to consider BG concentrations by using a jump diffusion model for meals and disturbances \cite{ccta}. The average meal rate in the jump part is chosen to be $3~\si{[Meals/Day]}$ between the hours 7:00 and 23:00 and $0.1~\si{[Meals/Day}]$ otherwise to consider that subjects eat less frequently at night. As for the diffusion part, a constant diffusion is added to the BG concentration state. The third model denoted as "Model 3" is the high fidelity model \cite{al2021glucose}. The meal times for Model 3 are drawn from uniform distributions as following:
$\mathcal{U}(6,8)~[\si{h}]$ for breakfast meals, $\mathcal{U}(12,14)~[\si{h}]$ for lunch meals, and $\mathcal{U}(19,20)~[\si{h}]$ for dinner meals.
% The meal times for Model 4 are allocated randomly (according to uniform distributions) within the following time intervals: 6:00-8:00 $[\si{h}]$ for breakfast, 12:00-14:00 $[\si{h}]$ for lunch, and 19:00-21:00 $[\si{h}]$ for dinner.
The carbohydrate intake for each meals is also drawn uniformly according to $\mathcal{U}(10,25)$ for breakfast, $\mathcal{U}(20,30)$ for lunch, and $\mathcal{U}(25,45)$ for dinner.
% with $\text{mean}\pm\text{SD}$ given by $45\pm10~\si{[g]}$ for breakfast, $75\pm10~\si{[g]}$ for lunch, and $85\pm 10~\si{[g]}$ for dinner.
We choose to simulate meals differently for Model 3 to test the strategies against a different type of stochastic disturbances.
Moreover, we consider an SMBG measurement error model \cite{ECCMEAS} for "Model 2" and "Model 3" as following
\begin{subequations}
    \label{eq:SMBG_model}
    \begin{align}
    \label{eq:SMBG_meas}
    y_{s}(k) &= x_{g}(k) + \sigma_{s}\left(x_{g}(k)\right)\varepsilon_{s}(k),\\
    \label{eq:SMBG_var}
    \sigma_{s}\left(x_{g}\right) &= \frac{1}{\kappa}\sigma_{2}\log\left(1+\mathrm{e}^{\kappa\left(x_g-4.2\right)}\right)+\sigma_{1},
    \end{align}
\end{subequations}
with $\sigma_{1}$ and $\sigma_{2}$ chosen in accordance to the ISO standard \cite{international2003vitro} to be $\sigma_{1}=0.415~[\si{mmol/L}]$ and $\sigma_{2}=0.1$, and $\kappa=5$. 
We did not add measurement noises to "Model 1" since the model is intended for a detailed comparison with the strategy in \cite{krishnamoorthy2020model} and we want to have the same model used in \cite{krishnamoorthy2020model} which did not consider measurement noises. Table \ref{tab:simulation_models} summarizes the models used for simulations in this paper.
\begin{table}[ht]
    \centering
        \caption{Glucose-insulin simulation models used in the paper}
    \label{tab:simulation_models}
    \begin{tabular}{|c|L{6.5cm}|}
\hline Model 1 & \footnotesize{Based on \cite{krishnamoorthy2020model}. Does not include a measurement noise model. Simulates FBG concentrations only. Includes process noise. Intended to be used for a detailed comparison with \cite{krishnamoorthy2020model} in Section \ref{sec:CompESC}}\\
\hline Model 2 & \footnotesize{Based on \cite{ccta}. Includes a measurement error model.} \\
\hline Model 3 & \footnotesize{Based on the model from \cite{al2021glucose}. Meals times and their sizes are drawn from uniform distributions. Includes a measurement error model. A diffusion term matching the one in \cite{ccta} is added to the state corresponding to BG concentration.} \\
\hline
\end{tabular}
\end{table}
\section{Results and Discussion}
\label{sec:Res}
In this section, we simulate our proposed strategy with different scenarios and compare it with three different strategies. The first strategy is the extremum seeking control strategy proposed in \cite{krishnamoorthy2020model} denoted as ESC\footnote{The sign of the gradient step was written to be positive in equation 6 in \cite{krishnamoorthy2020model} in a gradient \textit{decent} setup. Therefore, we used a negative sign instead since it is clearly a typo. Especially since the algorithm performed poorly when a positive sign is used.}. As for the second (denoted as 202) and third (denoted as Step) strategies, we use the standard of care titration strategies from \cite{kadowaki2017insulin} shown in Table \ref{tab:Standcare}. The 202 strategy adjusts the dose weekly based on the last day SMBG measurement while the Step strategy adjusts the dose weekly based on an average of the last three days SMBG measurements. 
\begin{table}[ht!]
    \centering
        \caption{Standard of care titration strategies.}
    \label{tab:Standcare}
\begin{tabular}{|l"c|c|}
\hline Strategy & SMBG $[\mathrm{mmol} / \mathrm{L}]$ & Dose adjustment $\Delta u[\mathrm{U}]$ \\
\thickhline \multirow{3}{*}{202} & $>6$ & $+2$ \\
& $4-6$ & No change \\
& $<3.9$ & $-2$ \\
\hline \multirow{4}{*}{Step} & $>9$ & $+8$ \\
& $8-8.9$ & $+6$ \\
& $7-7.9$ & $+4$ \\
& $5-6.9$ & $+2$ \\
& $3.9-4.9$ & No change\\
& $3.1-3.8$ & $-2$ \\
& $<3.1$ & $-4$ \\
\hline
\end{tabular}
    %\vspace{-4mm}
\end{table}
For our strategy, we simulate it with five different scenarios as following
\begin{itemize}
    \item \textbf{AdaOS}: Default strategy. Initial conditions $\hat{K}_{p}(0)=0.3$, $\hat{K}_{s}(0)=1$. A continuous score scale $y_{s}\in[0,H]$ is used with $H=10$.
    \item \textbf{AdaOS-H5}: Same as AdaOS but with a discrete score scale $y_{s}\in\{0,1,\dots,H\}$ with $H=5$.
    \item \textbf{AdaOS-F}: same as AdaOS but $\hat{K}_{s}=0$ (No PHG feedback) and $\hat{K}_{p}(0)=0.8$.
    \item \textbf{AdaOS-pf}: Same as AdaOS but subjects do not provide a PHG score on day $k$ with a probability $p_{f}$. If the subjects do not provide a score on day $k$, then $y_{s}(k)=y_{s}(k-1)$.
    \item \textbf{AdaOS-C}: Same as AdaOS-F and it is intended to be compared mainly with ESC (similar settings to ESC) in section \ref{sec:CompESC}. The reference is adjusted to be $r=5~\si{[mmol/L]}$ to match the one in ESC. The parameter $\hat{K}_{p}(0)$ is chosen to match the initial insulin dose for ESC in \cite{krishnamoorthy2020model}.
\end{itemize}
For all the scenarios, we let $\hat{\spsi}(0)=\left[0~0\right]^{\mathrm{T}}$, $P=\mathrm{I}$, $\lambda=0.9$, $\epsilon_{\phi}=10^{-3}$, and additive dithers on $\hat{K}_{p}(k)$ and $\hat{K}_{s}(k)$ chosen as $0.01\,\mathrm{square}(10k)$, with $\mathrm{square}(x)=\mathrm{sign}(\sin(x))$. Note that the choice of $\hat{K}_{p}(0)$ is important for the performance of the strategy. If it is chosen to be high, then the initial insulin doses would be high which can lower glucose concentrations too fast for the the estimation of $\hat{K}_{s}$ to catch up. This is especially due to the fact that $K_{s}$ has its main effect during the beginning of the titration phase. For our strategy, a value of $\hat{K}_{p}(0)=0.3$ gave us good results for all the simulations with the different models. For the case of AdaOS-F, there was no need to estimate $K_{s}$. Therefore, we chose $\hat{K}_{p}(0)=1$.\footnote{The code used for the simulations can be found on \url{https://gitlab.com/aau-adapt-t2d/T2D-AdaOS.git}.}
\subsection{Results with PHG}
In this section, we perform a one year simulation for 400 subjects with T2D. The first 200 subjects of the 400 were generated with Model 2, and the second 200 were generated with Model 3. For each subject, initial glucose and insulin concentrations were drawn uniformly together with parameters affecting insulin resistivity, insulin secretion, and the time constant for injected long-acting insulin. Table \ref{tab:Param_Model} summarizes the parameters drawn for each T2D model in addition to the parameters drawn for the PHG score model.
\begin{table}[h]
    \centering
        \caption{Parameters for generating subjects from Model 2 and Model 3. The state $x_{g}$ denotes the BG concentration while $x_{I}$ denotes the blood insulin concentration.}
    \label{tab:Param_Model}
    \begin{tabular}{|l|L{6.5cm}|}
        \hline
         Model 2 & \footnotesize{$x_{g}(0)\sim \mathcal{U}(13,20)~\si{[mmol/L]}$, $p_{4}\sim \mathcal{U}(0.5,2.5)$, $p_{7}\sim \mathcal{U}(0.5,2.5)$, $p_{1}\sim\mathcal{U}(1.5,2.5)$, $p_{6}$ and the initial conditions of the remaining states are calculated such that $x_{g}(0)$ is stationary. Diffusion $\sigma_{g}\sim \mathcal{U}(0.1,2)$.}  \\
        \hline
         Model 3 & \footnotesize{$x_{g}(0)\sim \mathcal{U}(13,20)~\si{[mmol/L]}$, $x_{I}(0)\sim\mathcal{U}(0.5,1)~\si{[mU/L]}$, $c_{1}\sim\mathcal{U}(0.01,0.03)$, $c_{2}\sim\mathcal{U}(1,2)$, $c_{4}\sim\mathcal{U}(1,2)$, and the initial conditions of the remaining states are calculated such that $x_{g}(0)$ and $I_{g}(0)$ are stationary. Diffusion $\sigma_{g}\sim \mathcal{U}(0.1,2)$.}\\
         \hline
         PHG & \footnotesize{$\rho\sim \mathcal{U}(2,20)$, $\bar{d}\sim \mathcal{U}(0.35,0.85)$, $h\sim \mathcal{U}{14,30}$, $\eta_{1}\sim\mathcal{U}(5,20)$. For AdaOS-pf, $p_{f}\sim\mathcal{U}(0.1,0.4)$.}\\
         \hline
    \end{tabular}
\end{table}
To compare the scenarios and the algorithms used in the simulations, we use the performance measures and their targets described in \cite{holt2021management} for glucose managements. The measures are shown in Table \ref{tab:GCMEAS}.
\begin{table}[ht!]
    \centering
    \vspace{0.2cm}
        \caption{Glucose management measures from \cite{holt2021management}. The unit for the ranges and glucose values is $\si{[mmol/L]}$.}
    \label{tab:GCMEAS}
    \begin{tabular}{|l|l|l|}
        \hline 
        Measure &  \% of time for BG in& Target\\ 
        \hline
         Time in Range (TIR) & $[3.9,10)$ & $>70\%$ \\
         \hline
         Time Above Range 1 (TAR1) & $[10,13.9)$ & $<25\%$ \\ 
         \hline
         Time Above Range 2 (TAR2) & $[13.9,\infty)$ & $<5\%$ \\
         \hline
         Time Below Range 1 (TBR1) & $[3,3.9)$ &  $<4\%$ \\
         \hline
         Time Below Range 2 (TBR2) & $[0,3)$ & $<1\%$ \\ 
         \hline
         Average Glucose (AG) & & $<8.6$\\
         \hline 
         Glucose Variability (GV)  &  & $<36\%$\\ 
         \hline 
         Glucose Managment Index (GMI)  &  & $<7\%$\\ 
         \hline
    \end{tabular}
\end{table}
\begin{figure*}[ht!]
    \centering
    \includegraphics[width=0.8\textwidth]{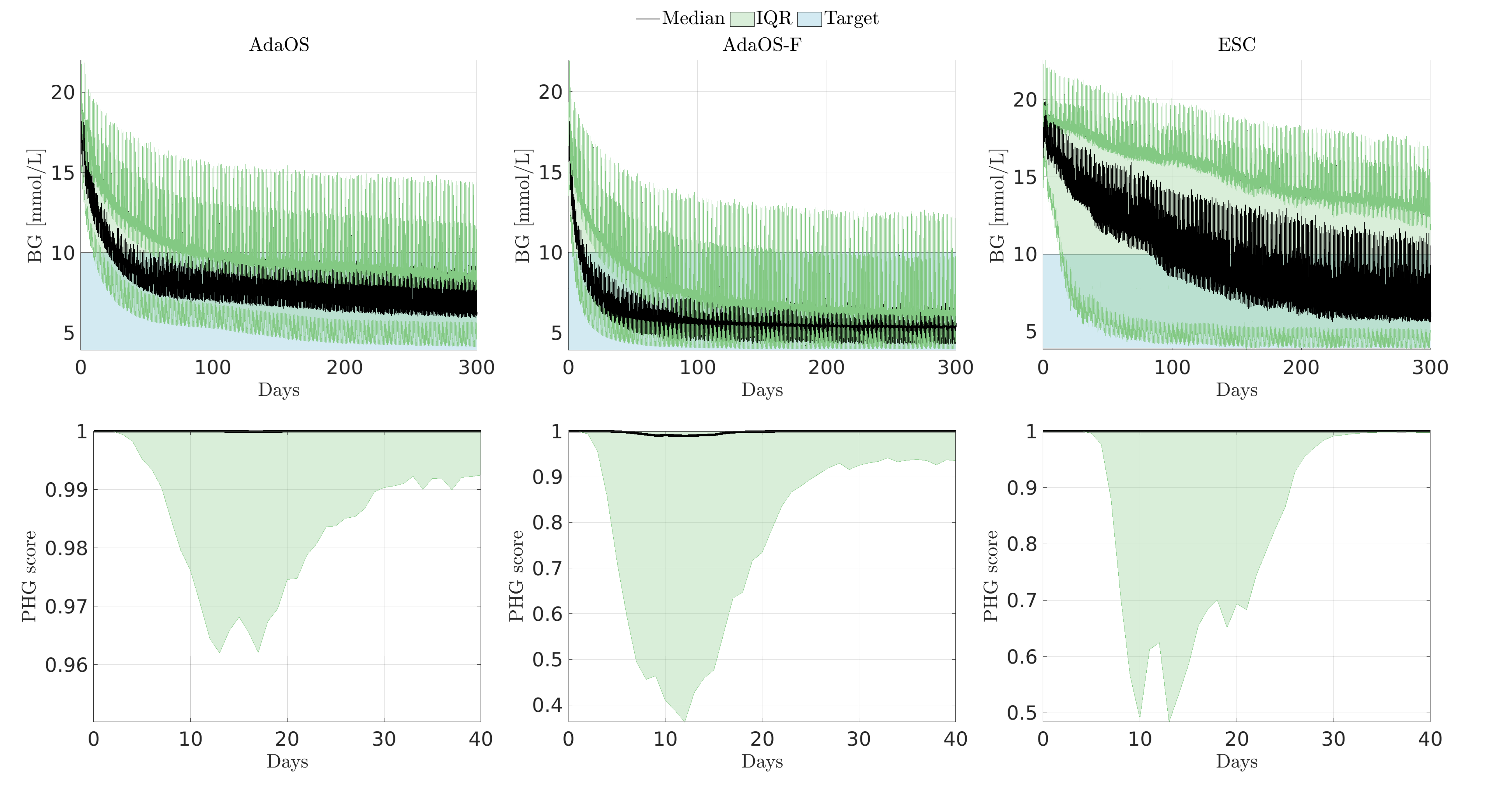}
    \caption{Simulation results for AdaOS, AdaOS-F, and ESC with Model 2 and Model 3. The time axis for PHG is only for 40 days since PHG is relevant during the initial titration phase.}
    \label{fig:mainfigure}
\end{figure*}
\begin{table*}[h]
        \centering
                \caption{Statistics for different scenarios and algorithms ({\color{red}red} numbers indicate values outside the target range).}
        \label{tab:MainResultTable}
        \begin{tabular}{|c"c|c|c|c|c|c|c|c|}
        \hline 
              & Mean TIR & IQR TIR & Mean TBR1 & IQR TBR1 & Mean TBR2 & IQR TBR2 & Mean AG & IQR AG\\
            \thickhline
             Target \cite{holt2021management} & $>70\%$ &  & $<4\%$ & & $<1\%$ & & $<8.6~\si{[mmol/L]}$ & \\
             \hline
             AdaOS & 95.35\% & 2.8\% & 1.2\% & $2\%$ & 0\% & 0\% & 8.43 & 4.77  \\
             \hline
             AdaOS-F & 96.77\% & 2.56\% & 2.11\% & $3.14$\% & 0\% & 0\% &  6.77 & 3.24 \\ 
             \hline
             AdaOS-H5 & 95.5\% & 3.06\% & 1.12\% & 1.76\% & 0\% & 0\% & 8.38 & 4.75 \\
             \hline 
             AdaOS-pf & 94.61\% & 3.03\% & 1.01\% & 1.5\% & 0\% & 0\% & 8.46 & 4.87 \\
             \hline
             Step & 91.08\% & 0.489\% & 2.3\% &  3.1\% & 0\% & 0\% & {\color{red}8.9} & 5.57\\
             \hline 
             202 & 77.96\% & 14.06\% & 3.39\% & 0.43\% & 0\% & 0\% & {\color{red}11.89} & 9.32 \\
             \hline 
             ESC & {\color{red}66.4\%} & 16.45\% & {\color{red}17.12\%} & 11.9\% & 0.83\% & 0.69\% & {\color{red}10.46} & 9.85\\
             \hline
             
        \end{tabular}
        \smallskip \\
        \begin{tabular}{|c"c|c|c|c|c|c|c|c|c|c|c|c|c|c|c|}
        \hline 
             & Mean TAR1 & IQR TAR1 & Mean TAR2 & IQR TAR2 & Mean Insulin & Mean GV & IQR GV & Mean GMI & IQR GMI\\
             \thickhline
             Target \cite{holt2021management} & $<25\%$ &  & $<5\%$ & & & $<36\%$ & & $<7\%$ & \\
             \hline
             AdaOS &  2.59\% & 1.81\% & 0.77\% & 0.97\% & 92.14 [U] & 25.5\%  & 8.22\% & 6.98\% &  2.1\% \\ 
             \hline
             AdaOS-F  & 0.85\% & 0.53\% & 0.27\% & 0.4\% & 162.96 [U] & 28.26\% & 11.38\%  & 6.25\%  & 1.41\% \\
             \hline
             AdaOS-H5 & 2.58\% & 1.84\% & 0.79\% & 1.04\% &  93.76 [U] & 25.92\% & 7.65\%  & 6.96\% & 2.06\% \\ 
             \hline
             AdaOS-pf & 3.59\% & 1.82\% & 0.8\% & 3.02\% & 92.64 [U] & $25.48\%$ & $8.05\%$ & $6.99\%$ & $2.11\%$ \\
             \hline
             Step &  4.8\% & 3.14\% & 1.78\% & 2.51\% & 125.55 [U] & 33.71\% & 7.29\% & {\color{red}7.18\%} & 2.42\% \\  
             \hline 
             202 &  14.99\% & 15.76\% & 3.66\% & 5.31\% &  57 [U] & 22.64\% & 21.91\% & {\color{red}8.48\%}  & 4.05\%\\
             \hline 
             ESC  & 4.6\% & 2.53\% & {\color{red}11.87\%} & 4.19\% & 69.61 [U] & 31.92\% & 25.86\% & {\color{red}7.86\%} & 4.28\%  \\
             \hline
        \end{tabular}
        \smallskip \\
        \begin{tabular}{|c"c|c|c|c|c|c|}
        \hline 
              & Mean $\text{PHG}_{>0.8}$ & IQR $\text{PHG}_{>0.8}$ & Mean $\text{PHG}_{<0.5}$ & IQR $\text{PHG}_{<0.5}$ & Mean $\text{PHG}_{<0.2}$ & IQR $\text{PHG}_{<0.2}$\\
             \thickhline
             AdaOS &  98.51\% & 0\% & 0.85\% & 0\% & 0.33\% & 0\% \\ 
             \hline
             AdaOS-F  & 89.26\% & 22.75\% & 6.06\% & 2.5\% & 3.1\% & 0\% \\
             \hline
             AdaOS-H5 & 98.4\% & 0\% & 0.89\% & 0\% &  0.49\%  & 0\% \\ 
             \hline
             AdaOS-pf & 98.38\% & 0\% & 0.88\% & 0\% & 0.54\% & 0\% \\
             \hline
             Step &  98.79\% & 0\% & 0.53\% & 0\% & 0.13\% &  0\% \\  
             \hline 
             202 &  99.93\% & 0\% & 0\% & 0\% &  0\% & 0\% \\
             \hline 
             ESC  & 87.7\% & 25\% & 8.86\% & 15\% & 5.86\% & 5\% \\
             \hline
        \end{tabular}
    \end{table*}
In addition to the measures in Table \ref{tab:GCMEAS}, we compute the mean long acting insulin dose, percentage of time for the PHG score $x_{s}$ being above 0.8 ($PHG_{>0.8}$), percentage of time for the PHG score $x_{s}$ being below 0.5 ($PHG_{<0.5}$), and the percentage of time for the PHG score $x_{s}$ being below 0.2. Note that we use $x_{s}$ here instead of $y_{s}$ since $x_{s}$ in \eqref{eq:PHGscore_day_k} represents the true score of how the subjects will rate their PHG symptoms and not the noisy (and possibly discrete) score $y_{s}$.
Table \ref{tab:MainResultTable} shows computed mean and Inter-Quartile Range (IQR) over the 400 simulations for each strategy or scenario. Additionally, Figure \ref{fig:mainfigure} shows the results for AdaOS, AdaOS-F, and ESC. From the results in Table \ref{tab:MainResultTable}, it can be seen that all the AdaOS variations have a mean satisfying the targets of the glucose management measures. AdaOS, AdaOS-H5 and AdaOS-pf have the best mean/IQR values for TIR, TBR1, TBR2, and for the PHG measures when compared to the other strategies. However, the mean GMI of AdaOS, AdaOS-H5, and AdaOS-pf is very close to the limit of its target range. AdaOS-F has better TIR statistics and mean/IQR values for AG, GMI, and GV when compared to the other strategies. However, AdaOS-F has a higher mean/IQR values for TBR1 when compared to the other AdaOS variation. Additionally, AdaOS-F preforms poorly for the PHG score when compared to the other AdaOS variations. This is expected since AdaOS-F is the version which does not use PHG scores as a feedback from the subjects. The Step strategy also performs as good as the AdaOS, AdaOS-H5, and AdaOS-pf in terms of PHG. However, the Step strategy does not perform as good as the AdaOS variations in terms of the glucose management measures and has mean AG and GMI violating their target. The 202 strategy has mean AG and GMI violating their target with poor performance for TIR, TBR1 and GV. Finally, ESC has mean TIR, TBR1, AG, TAR2, and GMI violating their targets. We provide a more detailed comparison with ESC in Section \ref{sec:CompESC}.
\subsection{Comparison with ESC}
\label{sec:CompESC}
\begin{figure}[ht!]
    \centering
    \includegraphics[width=0.48\textwidth]{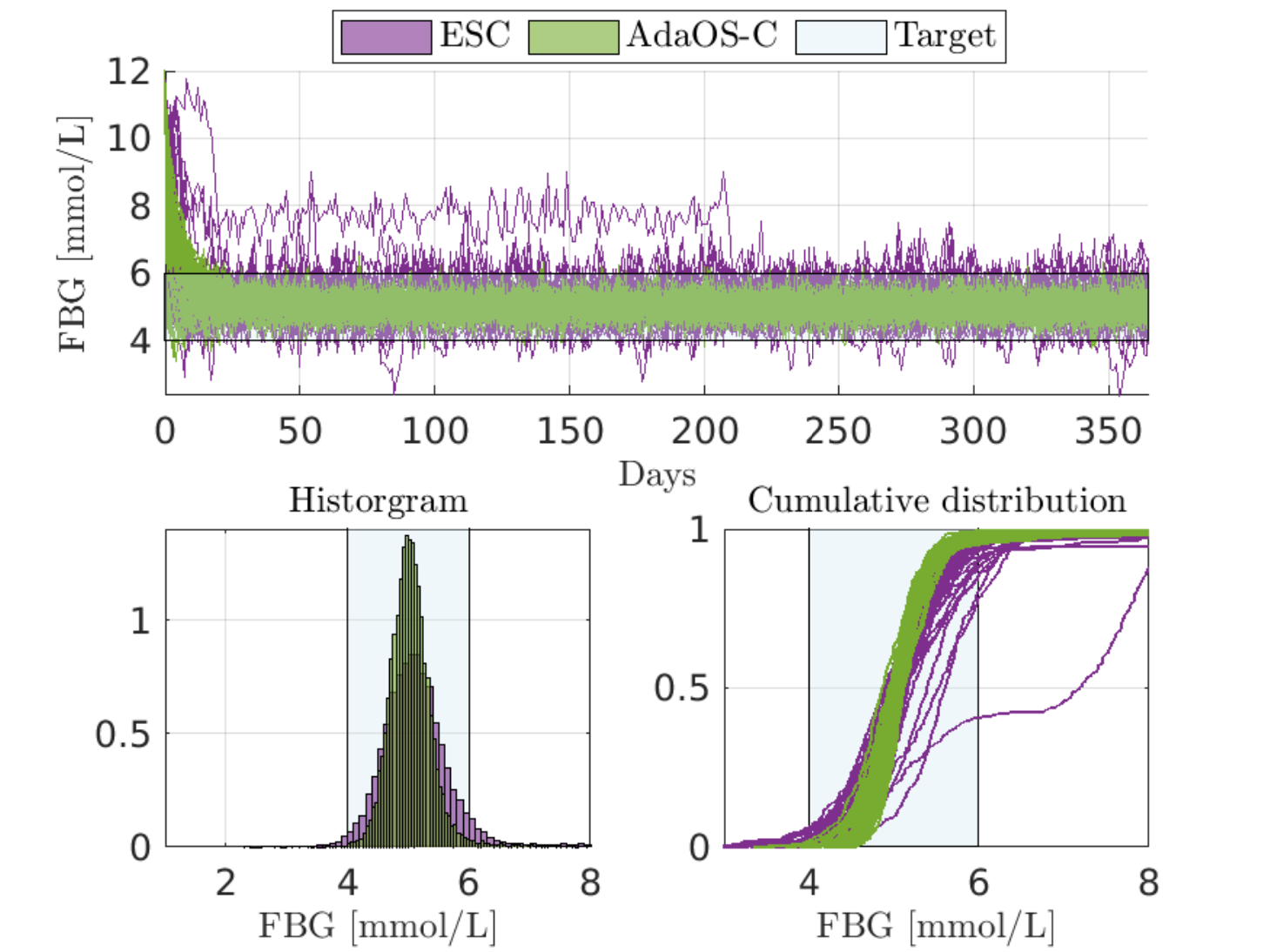}
    \caption{Simulating 61 T2D subjects using Model 1 with $p_{EGP}\in\{110:5:410\}$ for ESC and AdaOS-C}
    \label{fig:comparision}
\end{figure}

We compare ESC from \cite{krishnamoorthy2020model} with AdaOS-C using Model 1 without the PHG score since ESC does not account for it. We use simulation Model 1 with the same parameters used for the simulations in \cite{krishnamoorthy2020model} but with subjects having the parameter labeled $p_{EGP}\in\{110\!:\!5:\!410\}$ for each one of them. Thus, allowing $p_{EGP}$ to take larger values than the range in which \cite{krishnamoorthy2020model} tested their strategy against which was $p_{EGP}\in[350,380]$. The initial insulin dose for the simulated subjects in \cite{kristensen2004} was chosen to be $5~\si{[U]}$ with a fixed initial BG concentration of $12~\si{[mmol/L]}$. Therefore, we choose $\hat{K}(0)_{p}=\left(12-5\right)/5=1.4$ such that the initial insulin dose for AdaOS-C is also $5~\si{U}$. 
The results are shown in Figure \ref{fig:comparision}. In addition, we report in Table \ref{tab:AdaOS-SMBG_ESC} percentages of samples of FBG being within different ranges as done in \cite{krishnamoorthy2020model} with $[4,6]~[\si{mmol/L}]$ being the desired range, $[0,4)~[\si{mmol/L}]$ being the hypoglycemic range, and the $[0,3)~[\si{mmol/L}]$ being the severe hypoglycemic range.
\begin{table}[ht]
    \centering
        \caption{Average and worst case percentages of FBG samples within different ranges for each simulated subject with ESC and A$\si{da}$OS-C using Model 1. Unit for the ranges is $[\si{mmol/L}]$.}
    \label{tab:AdaOS-SMBG_ESC}
    \begin{tabular}{|c|c|c|c|}
        \hline
        \textbf{Average FBG} & $4-6$ & $<4$ & $<3$ \\
        \hline
        ESC & 92.61\% & 0.87\% & 0.03\%\\
        \hline
        \footnotesize{AdaOS-C} & 97.67\% & 0.085\% & 0\%\\
        \thickhline
        \textbf{Worst case FBG} & $4-6$ & $<4$ & $<3$ \\ 
        \hline 
        ESC & 40.44\% & 4.92\% & 0.55\% \\ 
        \hline
        \footnotesize{AdaOS-C} & 95.36\% & 1.37\% & 0\%\\
        \hline
    \end{tabular}
\end{table}
It can be seen from the results that the proposed strategy in this paper outperforms the ESC strategy and it is more robust to inter-subject variations. Note that the values which were chosen for $p_{EGP}$ in this simulation are realistic (see e.g. \cite{clausen2021new}). In addition, we point out that the maximum conditioning number for the covariance matrix of the RLS in ESC was $1.5\times10^{10}$ while the maximum condition number for the covariance matrix in AdaOS-C was $99.2$. The relatively high condition number in ESC when compared to AdaOS-C can be one of the reasons why AdaOS-C performs better. AdaOS-C ensures that the covariance matrix is well conditioned by using directional forgetting (the forgetting factors are not constant in the RLS) as discussed in Section \ref{sec:RLS}. 
\section{Conclusion and Future Work}
\label{sec:conc}
A model free approach based on an online stochastic optimization is proposed for insulin titration in T2D subjects. The proposed strategy combines the stochastic optimization algorithm AdaBelief with a RLS scheme to tune a feedback control law with SMBG measurements and personal feedback ratings from the subjects with respect to their PHG symptoms. Using simulations with different T2D models, the strategy was compared to different titration strategies from the literature with respect to the glucose management measures in \cite{holt2021management} and preventing PHG symptoms. The proposed strategy was shown to outperform the other titration strategies under different scenarios.
As two of the titration strategies were standard of care titration strategies, this indicates that the proposed strategy can be further developed to be implemented in a clinical setting. Furthermore, it shows the potential of including subjects' personal rating as feedback for automatic dosing strategies. Future work involves deriving theoretical guarantees for the proposed strategy, validating the strategy against other high fidelity T2D simulation models, testing different scenarios for PHG ratings, and to test it against a more accurate model for PHG when such a model become available.
\begingroup
%\raggedright
\bibliographystyle{IEEEtran}
\bibliography{literature}							% Litteraturlisten inkluderes
\endgroup

\end{document}